# Retrofitters, pragmatists and activists

## Public interest litigation for accountable automated decision-making

Henry Fraser[1] and Zahra Stardust[2]

*This paper examines the role of public interest litigation in promoting accountability for AI and automated decision-making (ADM) in Australia. Since ADM regulation faces political and geopolitical headwinds, effective governance will have to rely on the enforcement of existing laws. Drawing on interviews with Australian public interest litigators, technology policy activists, and technology law scholars, the paper positions public interest litigation as part of a larger ecosystem for transparency, accountability and justice with respect to ADM. The paper explores the tactics and strategies of what one participant described as "retrofitting" old laws to ADM. These go beyond creative legal argumentation, to encompass practices of community-building, collaboration on theories of change, canny selection of clients and causes of action, and aligning the interests of stakeholders in litigation. Naturally, the paper also contends with the limits of these strategies, and of the Australian legal system. Where limits are capable of being overcome, the paper presents findings on urgent needs: the enabling institutional arrangements without which effective litigation and accountability will falter. The paper is relevant to law and technology scholars; individuals and groups harmed by ADM; public interest litigators and technology lawyers; civil society and advocacy organisations; and policymakers.*

**Acknowledgements**

This research was funded by Australian Research Council grant number CE200100005.

We thank the peer reviewers for their thoughtful comments, which strengthened the paper.

We are grateful to Dr Alex Sinclair, Prof Chris Marsden, Prof Jeannie Paterson, Prof Paula O'Brien, Prof Christine Parker, and the attendees of the Monash-Deakin Digital Law Symposium and the ASSA Workshop on Strategic Public Interest Litigation for feedback on earlier drafts.

GenAI disclosure: The authors used Anthropic's Claude to do a final check for typographical errors, which we then corrected manually.

The authors declare no competing interests.

[1] Lecturer, Queensland University of Technology, School of Law; Associate Investigator, Centre for Automated Decision-Making and Society: h5.fraser@qut.edu.au

[2] Senior Lecturer, Queensland University of Technology, School of Communication; Associate Investigator, Centre for Automated Decision-Making and Society

# Introduction

SyRI, MiDAS, Robodebt – these are the names of unfair and poorly designed automated decision-making (ADM) systems deployed by government agencies around the world. Each was intended to detect overpayment of welfare benefits and to claw back money from welfare recipients. Each harmed thousands of people already experiencing marginalisation. Each was implemented with apparent disregard for the law and a distressing lack of empathy. Despite efforts at other forms of accountability, the intervention which most dramatically exposed the injustice of these systems and precipitated change was litigation (*NJCM v The Dutch State* 2020; *Zynda v. Zimmer* 2015; *Cahoo v. SAS Analytics Inc* 2019; *Prygodicz v Commonwealth of Australia (No 2)* 2021). For example, the Robodebt class action against the Australian government led to a settlement where the government agreed to repay over $751 million to wrongly targeted welfare recipients, and to withdraw over $1 billion worth of improperly issued debt notices (*Prygodicz v Commonwealth of Australia (No 2)* 2021). Where a series of administrative tribunal decisions, news reports, a Commonwealth Ombudsman's report and a senate committee report seemingly could not generate accountability for the Robodebt scandal (Carney 2019), the threat of liability in negligence and unjust enrichment ultimately moved the needle. Similarly, in the US, a series of strategic lawsuits concerning harms to young people from "AI companion" chatbots successfully brought risks from chatbots to the attention of the public and prompted a rapid and decisive regulatory response in New York, California, Australia, and several other jurisdictions (Fraser, Ciriello, and Szczuka 2026). These cases used both private and public law to challenge government and corporate uses of ADM technologies.

If litigation has been a key catalyst for accountability in relation to ADM, why hasn't there been more (and more successful) litigation? What strategies and tactics might generate greater success? What conditions and resources are most needed to enable effective public interest litigation? And where does public interest litigation run up against hard limits? Motivated by both curiosity and urgency, we interviewed public interest litigators, technology policy advocates, and technology law scholars in Australia, a country with an explicit policy of using existing law to manage harms from ADM (*National AI Plan* 2025). We asked about the kinds of ADM harms they were seeing; opportunities for impactful public interest litigation; pros and cons of various legal doctrines; the relationship between litigation and broader advocacy movements; the barriers to corporate and government accountability; and the enabling conditions that make public interest litigation successful for ADM harms.

Our article is divided into four sections. We begin by identifying the *roles and goals* of public interest litigation with respect to ADM. Given that public interest litigation can only address narrow questions of law, it is essential for it to be embedded in movement building, as one tool in a larger toolkit not only for law reform but system change. We then set out key *strategies* and *tactics* of public interest litigators in challenging ADM-related harms, using the concept of "retrofitting" – introduced by one participant - to organise our findings. Retrofitting involves adapting old laws to new harms, selecting doctrines, clients and methods which take best advantage of the law's capabilities. We then document some key *limits* of public interest litigation, especially its difficulties in grappling with collective, structural, automated harms which require more systemic, political solutions. We finish by documenting fundamental *enablers* for effective public interest litigation about ADM including meaningful transparency about ADM across the lifecycle of accountability processes, better funding of public interest litigation and the civil society ecosystem, and building of networks and institutions to share information and mobilise action.

In doing so we make three key conceptual contributions to the literature on public interest litigation and ADM. Firstly, we show that in a terrain where the law is uncertain, retrofitting existing laws for ADM harms means more than just adapting doctrine. It encompasses a range of practices, choices, and trade-offs in developing, bringing and evidencing suits, and mobilizing the resources to do so. Secondly, we argue that navigating alignments and misalignments of interests among stakeholders – including AI developers and deployers, affected communities, litigants, their advocates, the state and the public – is a key part of the craft of retrofitting. Thirdly, we put forward the concept of the "transparency-accountability lifecycle". Building on scholarship on AI opacity (Selbst and Barocas 2018; Burrell 2016), we explain how different kinds of transparency matter at different times to different stakeholders, and how transparency is both a condition for and a product of litigation.

# Background

## The state of the ADM accountability ecosystem in Australia

Where ADM is used to make consequential decisions it can cause harms of all kinds (O'Neil 2016; Buolamwini and Gebru 2018; Yeung 2019). This is especially true for "machine learning" and "artificial intelligence" (AI) technologies. "AI" describes computational systems that automate complex information processing tasks through algorithms, statistical methods, and data-driven optimization techniques, including "machine learning". "Machine learning" is the iterative process of automatically adjusting the parameters of an algorithm through exposure to data to optimize performance on specific tasks. Bender and Hanna have humorously proposed a less grandiose name: SALAMI (Systemic Approaches to Learning Algorithms and Machine Inference) (Bender and Hanna 2025, 5).

Regardless of how we name them, such tools generate life-changing decisions at unprecedented scale and speed. While the government misuses of ADM described above were particularly egregious, many serious harms have come to light over the past 5 years in both the public and private sectors. Misuses of ADM in law enforcement (Angwin et al. 2016; Sentas and Pandolfini 2017); autonomous vehicle accidents (Fraser, Simcock, and Snoswell 2022); toxic AI chatbots (Fraser and Suzor 2025; Roose 2024; Xiang 2023); facial recognition and intrusive surveillance in retail settings (Jung and Kwon 2024; Kind 2025; *Commissioner Initiated Investigation into Bunnings Group Ltd (Privacy)* 2024); denial of financial services to sex workers and sexual health providers (Stardust et al. 2023); intrusive social scoring (Blakkarly 2022); and discriminatory decision-making in education (Hern 2020) and finance (Garcia, Garcia, and Rigobon 2024), have all made news.

There is a pressing need for scholarship on how to apply existing laws to ADM harms. The Australian government recently abandoned a proposal to introduce "mandatory guardrails" for high-impact AI (*Safe and Responsible AI in Australia* 2024), in favour of a National AI Plan which explicitly states that "established laws remain the foundation for addressing and mitigating AI-related risks" (*National AI Plan* 2025). Even if the appetite for regulation were stronger, emerging AI and ADM regulation has generally been constrained by a "risk-based" approach and only applies to a relatively narrow range of high-risk applications (Mahler 2020; Schuett 2023).

Existing law therefore has an important role to play in regulating ADM. But the application of existing law to new technologies is uncertain (Brownsword 2008; Collingridge 1980; Bennett Moses 2011). That uncertainty is particularly pronounced for ADM technologies which have developed very quickly over the past decade. Absent regulation, law develops through litigation – including "public interest litigation".

A precise definition of public interest litigation is elusive (McGrath 2008; *Oshlack v Richmond River Council* 1998; Marshall and Hale 2014; Taylor 2020). One of its key features, according to the Australian Law Reform Commission, is that it resolves a question of law, develops the law and addresses matters in which the broader public has an interest (*Costs Shifting - Who Pays for Litigation?* 1995, pt. 13.2). Ramsen and Gledhill identify four main features of "strategic litigation":

1. it seeks outcomes with a long-term impact, going beyond the origins of the claimant's complaint;
2. it is adaptable to a range of purposes;
3. its objectives are multi-faceted and go beyond creating effects within the court system; and
4. it views "litigation" broadly, to include tribunals and international mechanisms of redress (Ramsden and Gledhill 2019).

It is generally recognised that public interest litigation is an important vehicle for the protection of the rights of marginalised or disadvantaged groups, or for driving social or legal change in the absence of political will (Durbach et al. 2013). Class actions, which were introduced into Australia in 1992 under Part IVA of the Federal Court of Australia Act 1976 (Cth) (with state regimes also allowing class actions in state courts), provide one avenue for groups to pool resources in relation to shared harms.

There are numerous stakeholders in public interest litigation, including ADM developers and deployers, individuals and communities affected by ADM, litigants (parties in a suit), litigators (their lawyers), civil society organisations, the state, and the general public. Although the public and litigants are the most impacted by the outcome, it is not necessarily these groups who drive the process. Scholars have explored differences between pro bono, cause and movement lawyering – with differences mostly arising from the role played by lawyers relative to other stakeholders in litigation (Meeusen 2026). Our interviews indicated that, in Australia, public interest litigation is often propelled by professionals: consumer groups and lawyers who have already built case theories, who are looking to "recruit" litigants to put theory into practice. This fits the profile of "cause" lawyering, though many participants emphasised the importance of ensuring more voice for other stakeholders, in line with "movement" lawyering.

Public interest litigation in Australia faces several well-known barriers. Costs of litigation may be prohibitive (Edwards 1999; Ginnivan 2016). Losing parties may, in addition to their own legal costs, be required to pay the costs of counterparties under "adverse costs orders" (*Costs Shifting - Who Pays for Litigation?* 1995). Finally, those most motivated to bring claims may be obstructed by a lack of "standing" or eligibility (Cane 1999). Even so, public interest litigation has influenced some of Australia's most significant socio-political issues, including the recognition of Indigenous land rights (*Mabo v Queensland (No 2)* 1992), environmental protection (*Tasmania Dams Case* 1983), asylum seeker detention (*Al-Kateb v Godwin* 2004), and same-sex marriage (*Commonwealth of Australia v Australian Capital Territory* 2013).

By contrast, litigation has yet to produce significant legal clarification with respect to ADM, although Robodebt litigation certainly precipitated political action. Few significant cases dealing with ADM have come before the court, and the success of these cases has been mixed. The *Pintarich* case – where the majority held that a computer-generated letter from the tax office was not a "decision" capable of judicial review because it was not produced by a human mental process of deliberation – generated more confusion than clarity, despite Justice Kerr's powerful dissent. In the Robodebt litigation, an action for judicial review led to consent orders stating that the government's approach to raising automated welfare debts (comparing actual monthly income to average monthly income) was irrational (*Amato* 2019). Consent orders do not, however, produce legal precedent.

As for litigation in private law, a subsequent Robodebt suit for negligence and unjust enrichment concluded in a settlement, which also does not produce binding precedent (*Prygodicz v Commonwealth of Australia (No 2)* 2021). More promising, the Australian Competition and Consumer Commission's consumer lawsuit against the travel website, Trivago, successfully applied the law of misleading and deceptive conduct in relation to claims about the operation of a ranking algorithm (*Australian Competition and Consumer Commission (ACCC) v Trivago NV* 2020). The case did not settle any particularly thorny issue of law, though it did establish a helpful roadmap for leading expert evidence about algorithmic systems and behavioral science (Fraser, Simcock, and Snoswell 2022).There is much more scope for public interest litigation to tackle the problem of ADM accountability.

## Aims and method

This project uses applied qualitative social research to investigate the challenges, opportunities, barriers and enablers for effective public interest litigation for ADM accountability. Applied social research addresses problems raised by organisations, governments or businesses and is designed to solve immediate and persistent problems (Hall 2008). The project was conducted within the Centre for Automated Decision-Making and Society, which was specifically funded by the Australian Research Council as a 7-year 'Centre of Excellence' initiative, bringing together universities, industry, government and community to investigate the development of responsible, ethical and inclusive ADM. The goal of this project was to achieve real-world change, bringing together stakeholders who encounter and challenge ADM systems in their daily practice, and synthesizing their insights into practical recommendations that can be employed by advocates, communities, and practitioners.

To do so, we recruited and interviewed twenty candidates with experience in public-interest litigation, advocacy, technology law and policy. We obtained ethics approval for the study from our faculty ethics committee before commencing. Participants included barristers, solicitors, policy advisors, academics and consumer advocates from both public and private institutions. We used our own networks, based on our previous experience as lawyers and law reform advocates, and developed during our research-fellowships at the Centre for Automated Decision-Making and Society to recruit participants initially, and then used snowball sampling based on participant recommendations. This sample size is typical for qualitative research, which seeks to elicit rich, in-depth, textural data that offers insights into the world, rather than generating population-level or generalisable data. Our sample size was sufficient for us to reach saturation point, whereby new data was not yielding additional insights (Hennink and Kaiser 2019). It is also worth noting that the Australian public interest litigation and digital rights ecosystem is small compared with jurisdictions like the US or UK.

We conducted and recorded one-hour semi-structured interviews and gave participants the opportunity to review and edit transcripts for clarity and confidentiality. We used an outline of questions for the interview, but allowed conversation to flow naturally based on participants' answers, interests and expertise. The interviews were specifically related to public interest litigation in the context of ADM. ADM provides a unique challenge for public interest litigation, given the speed of development of ADM technologies, their application to large domains of economic and social life, the lack of accountability of governments and corporations for harmful ADM, uncertainty about how the law applies, and the sheer scale of actual and potential harm.

We then conducted qualitative thematic analysis of the data (Braun and Clarke 2006), identifying, analysing and interpreting patterns of meaning (themes). Our first set of codes were organised around the structure of the interviews. These included opportunities for public interest litigation (identifying relevant cases and areas of practice); barriers (which ranged from aggressive litigation and system opacity to lack of access to justice and systemic harms); and enablers (including funding, aggregating institutions, law reform, legal cultures, referral networks, and monitoring). We then developed further conceptual codes, taking an iterative approach to coding that drew from our contemporaneous file notes, reflections and memos on the significance of the data (Strauss and Corbin 1998). These codes included doctrines, power and movement lawyering – featuring data about who drives litigation, the limits of the law, and the role of lawyers – as well as retrofitting, aligning stakeholder interests, and the transparency-accountability lifecycle.

There are several limitations to our study. Participants were mostly professionals (lawyers, members of civil society organisations, academics).[3] Our understanding of the experience of litigants was thus mediated by these professional advocates, notwithstanding their deep knowledge of the issues. A key direction for future research is prioritising perspectives from impacted communities who are experts in their own experience and can offer bottom-up insights into how ADM operates. Only one participant was an administrative law expert. Consequently, although our study was partly motivated by the Robodebt cases (of which one was a judicial review action), interviews did not generally produce deep dives into ADM and administrative law.

The data is most relevant to the Australian legal context, where doctrines like administrative law and discrimination are idiosyncratic, and where the digital rights advocacy ecosystem is relatively small. Other jurisdictions have different institutional parameters and legislative contexts (for example, in relation to costs waivers). However, our findings around tactics of retrofitting may be relevant to other common law jurisdictions, especially those that have not introduced strong AI regulation.

[3] One of our interviewees, however, was personally impacted by Robodebt and was a litigant in the class action.

## Related scholarship

This article builds on and contributes to the scholarship of how to develop and apply existing laws (as well as new regulations) to ADM. Scholars have analysed both conceptual and practical challenges of doing so. Important work, for instance, explores whether and how foundational concepts such as "accountability", "public interest" or "explanation" apply to decisions made or mediated by ADM (Cobbe 2019; Miller 2019; Singh, Cobbe, and Norval 2018; Cohen and Suzor 2024; Wachter, Mittelstadt, and Russell 2018). Other work documents the ways in which the opacity, complexity and inscrutability of AI systems obstruct accountability (Burrell 2016; Pasquale 2015; Selbst and Barocas 2018). A growing literature critically evaluates the effectiveness of new regulations such as Europe's AI Act (Veale and Borgesius 2021; Schuett 2023; Mahler 2020; Engler and Renda 2022; Fraser et al. 2024; Hacker and Holweg 2025), and existing doctrines - such as equity (Zhou et al. 2024; Balkin 2015), discrimination law (Sheard 2022; Barocas and Selbst 2016; Mann and Matzner 2019), tort (Karnow 2016; Yoshikawa 2018; Selbst 2020; Fraser and Suzor 2025; Buiten, de Streel, and Peitz 2021) and administrative law (Ng 2025; Cobbe 2019; McGurk and Tomlinson 2025) – in addressing harm and injustice from ADM.

We extend this literature by exploring the practical and tactical implications, for litigants and litigators, of the strengths and weaknesses of different doctrines. In doing so, we connect the conceptual literature to more practically oriented scholarship. Private law scholars have delved into challenges and opportunities of developing and leading evidence and civil procedure in cases involving AI and ADM (Caruso, Legg, and Phoustanis 2019; Llorca et al. 2023; Fraser, Simcock, and Snoswell 2022). Scholars have highlighted institutional practices and attitudes in government which obfuscate ADM harms, and obstruct accountability (Carney 2019; Henman 2021, 2025). And public law scholarship has tackled the challenge of designing reviewable ADM systems (Cobbe, Lee, and Singh 2021), evidencing public law errors in ADM in judicial review (Tomlinson, Sheridan, and Harkens 2020) and using data science to formulate new approaches to reviewing the rationality of ADM (Simcock 2026).

We also align with literature theorising strategies and tactics for public interest litigation such as "winning through losing" (NeJaime 2010), the relationship between law and organising (Cummings 2009), and the role of public interest litigation in raising awareness and generating political pressure (Harlow and Rawlings 2013).

Our contribution to that practical literature is not only to bring theories of litigation to bear on the issue of ADM accountability, but also, through our qualitative method, to add granular detail on the practical craft of public interest litigation in the Australian context. Rather than doing historical, political or doctrinal analysis of cases and movements (like NeJaime or Cummings), our findings come from thematic analysis of our conversations with Australian domain experts. For instance, we enrich Tomlinson et al.'s prognostications about practices of fact-finding and expert evidence likely to evolve to deal with ADM (Tomlinson, Sheridan, and Harkens 2020) with details about how participants approach decisions regarding cost, resources, and politics. Our approach thus contributes rich insights about how power, politics and practice affect ADM and public interest litigation in Australia.

# Roles, goals and strategies of public interest litigation

Public interest litigation may not always produce systemic change on its own. Indeed, our interviews suggested that it is most impactful when nested in broader strategies of advocacy. Public interest litigation can help build momentum for social movements, but it also depends on movement-building to develop the capabilities, resources and knowledge to bring litigation in the first place. We set out the basis for these findings below.

## Just one tool in a larger toolkit for justice

Almost all participants agreed that public interest litigation has *some* role to play in ADM governance, though their levels of enthusiasm varied. Participants with a background in litigation tended to be the most positive. For example, Lizzie O'Shea, a solicitor specialising in public interest litigation, working in a large Australian plaintiff law firm, and the chair of advocacy organisation, Digital Rights Watch, said:

> regulatory action by a regulator, or also regulatory reform, I don't think alone those things are sufficient to deal with the challenges presented by new technologies, in my experience. I do think there's a role for private enforcement of rights, as in people, litigants, either alone or as part of a group, bringing actions.

The framing of litigation as a kind of private regulation (echoed by a number of participants) represents an optimistic view: that individuals enforcing their rights in court using existing laws could exert a broader regulatory effect. This view aligns with a body of regulatory theory literature with a law and economics flavour, which views litigation as a form of user-pays, private regulation which may be more cost effective and efficient than regulation under certain conditions (Luff 2011; Posner 2010). One such condition is when change outpaces law-making – and that is very much characteristic of ADM.

Participants with other backgrounds were less enthusiastic. Academics, such as Professor David Lindsay of UTS and Professor Lyria Bennet-Moses of UNSW were interested in "zooming out", and critically evaluating the whole regulatory architecture for governing ADM. As Lindsay put it,

> public interest litigation and class actions are necessarily looking at things after the fact, after there have been harms. But the ideal is to try to prevent harms before they arise.

Lindsay and others saw litigation as capable only of achieving "incremental changes"; with regulatory intervention needed for more systematic or significant reform.

Informed by his experience of government decision-making, Professor Terry Carney, a longtime member of the Administrative Appeals Tribunal, and an expert in social security and administrative law, was pessimistic for other reasons. He pointed out that litigation in relation to government decision-making is limited, as courts are only permitted to review the legality, rather than the merits, of decisions of the executive branch of government. He suggested that the development of better means to manage merits review of administrative action would permit accountability across a much wider range of concerns. This is perhaps a good reason to understand public interest litigation more broadly, to encompass this kind of administrative action, as well as lawsuits in court (Ramsden and Gledhill 2019).

Nonetheless, while many participants preferred other avenues for ADM accountability, very few totally dismissed public interest litigation. Vicki Sentas, a legal scholar, warned about the potential for litigation to have "perverse effects" – prompting police, for example, to simply tweak their ADM systems to cement, rather than remedy, injustice. Despite her reservations she remained heavily involved in public interest litigation through coordinating the Police Powers Clinic in collaboration with UNSW and Redfern Legal Centre. Most emblematic of the less enthusiastic participants was Kate Bower, who at the time of the interviews worked at the consumer peak body, CHOICE, and who is now working at the Office of the Australian Information Commissioner. For her, litigation was "the last stop" when all other forms of advocacy fail. For another public interest litigator turned legal scholar who preferred to remain anonymous, litigation was something that was needed "in the meantime" while waiting for more systematic regulation. These participants perceived litigation to be less appealing than other forms of advocacy and regulation, but still acknowledged that it had a place within a broader ecosystem of accountability.

## From law reform to systemic change

If public interest litigation does have *some* role in promoting ADM accountability, what is that role? Our participants presented a range of views on the goals of public interest litigation which were broadly consistent with Ramsden and Gledhill's pluralistic characterization of "strategic" litigation (Ramsden and Gledhill 2019). Min Guo, a barrister who represented the plaintiffs in the Robodebt class action, summed up the goal of public interest litigation as "systemic change", and this view was common among participants. Interestingly, only one participant included direct pro bono work – lawyers helping individuals or organisations gratis without necessarily adopting a broader political framing (Meeusen 2026) – in their descriptions of public interest litigation. This may have been a result of participant selection. That participant was the only lawyer from a large commercial firm (as opposed to a plaintiff firm). In any case, participants presented different overall strategies for achieving such change and our analysis identified three main approaches: asserting the law, developing the law, and advocating change by combining litigation with other forms of advocacy.

One basic strategy is to use public interest litigation to remind ADM developers and deployers that they are not above the law. Professor Ed Santow, a former head of the Public Interest Advocacy Centre (now the Justice & Equity Centre), former Australian Human Rights Commissioner, and now a Professor at UTS said of public interest litigation:

> it can show, in a real, live case, that there is an applicable law, it can be applied, it has been applied… [T]hat will then be the foundation for people who are in the same position as the initial plaintiff to then be able to go, well, we can take this to court if we need to, but let's all save ourselves a lot of time, effort, and trauma, and just apply the law as the court has determinatively set out.

Michael Rivette, a senior barrister, saw legal judgments in litigation as setting new "boundaries" in companies' appreciation of risk. Companies recognise that if they cross that boundary, they are in a "risk area", he said. The strategy in such cases may therefore be to achieve systemic change through a change in *perception* of the law rather than a dramatic change in the law itself.

Another strategy was to develop the law in the public interest through the enforcement of a private interest. Here is how Min Guo described this strategy:

> [T]o bring matters before a court that will, firstly, protect or give an outcome to an individual or a corporation maybe who has been wronged. And secondly to pitch those cases in a broader context… in an attempt to give us new law.

This strategy fits within a broader theme we observed that public interest litigation is an exercise in aligning interests and incentives: in this case, aligning private interests in disputes with broader public interests.

In other circumstances, the goal might be to draw attention to a problem with existing law and policy, in order to prompt changes in policy and legislation. Describing the work of the Public Interest Advocacy Centre (PIAC), Principal Solicitor (and now CEO of Redfern Legal Centre) Camilla Pandolfini said:

> [W]e aim to change policies, practices, legislation that we think is unfair through a combination of strategic litigation and advocacy. We combine advocacy policy with strategic litigation and we try to think about what is the best way to bring about change in this particular area.

Where the goal is systemic change, public interest litigation works best when it is aligned with a broader strategy of activism. When public interest litigation is integrated into strategic advocacy, its success need not depend on "winning" court cases. Lizzie O'Shea explained that even when public interest litigation, "doesn't work, it can be really effective." She gave the example of a suit about harassment of staff and patients at the front of the Fertility Control Clinic in Melbourne. The Clinic sued the city council for failing to prevent that harassment. She described the outcome in these terms:

> [W]e lost the case. But then what they did was, it cleared the path for regulatory reform, because every time you ask for the introduction of safe access zones, parliamentarians would say, oh no, the local council's responsible, you've just got to get them to act on it. And then we sued them and found out they didn't have a duty to act. So then, okay, well, you need to fix this in a regulatory way, you need to implement reform. So, there's way in which you can lose, which then becomes a platform for change and that can be very effective.

What O'Shea describes aligns closely with what NeJaime describes as "winning through losing" – in this instance, using the litigation loss to galvanise and mobilise activists, and appeal to the public and other non-judicial actors for change (NeJaime 2010).

Several participants took similar views, with Min Guo suggesting that this strategy amounts to demonstrating to the public that "The law is not on our side, therefore the law needs to be changed." They explained that galvanising public opinion in this way requires a combined approach to advocacy integrating legal action, media action, policy advocacy and awareness raising. These reflections are consistent with research by Newhouse and others (who have combined litigation with coronial inquests, complaints and education) suggesting public interest litigation is "most effective when conducted in conjunction with public advocacy within a multi-faceted campaign for change" (Newhouse et al. 2023).

## Embedding litigation in advocacy and movement-building

Participants explained that the success of the "combined" approach to public interest litigation described above depends on it being deeply connected to social movements, broad networks and communities. Some explained their approach to building these communities. Rivette, for example, spoke about a "brains trust" which included both academics and lawyers who could help him to keep abreast of international developments in privacy law which he might potentially import. "[T]here's a great depth of knowledge," he said, "and I see myself as a compiler of that".

A human rights lawyer who worked with Indigenous communities favoured an approach to networking and knowledge-sharing that was still more structured and forward-looking:

> Well, what I do, and what we have done in the past is to bring people together. We call it a "hackathon", to "hack" particular social issues or problems, to come up with scenarios and legal theories to address them, and then share the information around. Sort of like calling a session, or a symposium, where people meet, with the particular intention of developing new legal strategies. This would require industry participants, people who understand the dark web, people who can bring data upon which the case theory can be based… You bring all the key players together with lawyers, and you develop a strategy… You must prepare for these events. You survey stakeholders to deliver a list of the top issues that have potential for litigation and then workshop them. And then you take the best of the solutions with the intention that individuals or organisations can run with any of them. I think sharing information, working together is really important.

By developing theories of how the law could develop to deal with likely scenarios, litigators then need only wait for (or may even actively seek) a client whose circumstances are sufficiently close to the theory. They can then act quickly, decisively, and with the confidence that their approach – which must address the complex and cross-disciplinary nature of ADM systems – is supported by a rigorous and collaborative process of analysis and peer review.

# Creative tactics of public interest litigators

What kinds of tactics are useful in bringing public interest litigation? And how can litigators calibrate these to ADM harms? In dealing with harms from new technologies, the task of the public interest litigator is often to "retrofit" old laws to new harms – this is necessary in the absence of general rights or specific regulation of ADM. But retrofitting is not merely a doctrinal exercise. It entails a whole range of practices and tactics. While participants mentioned many tactics, we noticed a unifying theme of *alignment*. Retrofitting tended to serve an overarching goal not only of aligning the law with new circumstances, but of aligning interests of diverse stakeholders, including individual and class action litigants, groups affected by ADM, litigators, funders and, of course, the broader public.

## Retrofitting old laws to new harms

Many of our participants recognised that our existing laws do not neatly "fit" the types of wrongs and harms emerging from the increasing use of ADM in consequential contexts. This is particularly the case in Australia, which lacks a federal Bill of Rights. Some states have charters of rights (bringing more potential to challenge ADM harms) but generally litigants must rely on an uneven patchwork of rights (Brennan et al., 2009). Where harms are not specifically covered by legislation, litigants have to rely on adapting common law or old legislative remedies to new circumstances. Min Guo described this as a process of "retrofitting" existing laws. He explained the challenge of legal retrofitting:

> The problem is that any litigation, at least in the common law system, is going to have to be through the constraints of the law as it is now. The constraints of judge-made law are shaped very heavily by precedent or, for that matter, what parliament says is the law. And so – particularly in a field like litigation where you're targeting decisions that are being made by computer – you're always trying to (because it's the only way to do it) find a way to retrofit existing causes of action, which were conceived of by judges centuries ago, or laws that were written in the previous decade with a different purpose, because they're the only tools in your armoury. You have to find a way to use them to address a new world problem.

An experienced litigator from a legal advocacy organisation gave several examples of legal retrofitting in non-ADM contexts. In one case he used property law to protect the interest of the individual in preventing the arbitrary government confiscation of phones in immigration detention: "When Peter Dutton decided to take peoples' phones off them in immigration detention, there was no human right that we could rely on to argue that detainees should keep them. We relied on the fact that the government is not allowed to take your property without a proper statutory basis. We had to rely on property law. Property law trumps human rights." He thought that such an approach – falling back on basic flexible, common law doctrines – would fit well for the new kinds of harms associated with ADM.

All cases in common law jurisdictions to some extent involve applying old law to new circumstances, but the need for retrofitting is especially salient for ADM. It is a fast-moving technology. It causes new kinds of harms including automated denial of access to critical private and public services, collective harms like automated bias; and harms which do not fit usual patterns of responsibility, such as autonomous vehicle accidents.

## Choosing the right action, client and remedy

Retrofitting is not simply about making creative arguments in court. Participants were highly tactical about formulating and bringing cases. One of the first choices that litigators face when formulating a claim about ADM is to choose a cause of action. In Lizzie O'Shea's words: "what kind of harm are we going to seek to remedy and what does that remedy look like?" These constraints are not merely doctrinal (what harms can the law remedy?) but also practical. Public interest litigators can be tactical about which clients and cases to take on - as Min Guo put it, "Is it a set of facts that actually lends itself nicely to existing law?" This question matters when litigators and activists have scarce resources to apply toward their strategic goals. One community litigator explained, "We don't have the resources to run many cases. So the cases that we take on must have a strategic focus".

Participants also pointed out a range of other tactical questions: including available remedies and the amount of work needed to update the law. It makes sense for litigators to retrofit causes of action that require less legal and conceptual heavy lifting. Where litigants have already had success with a particular doctrine, it will generally be efficient to continue to use (and further develop) that cause of action. Megan Richardson, a privacy and technology law scholar, spoke of her long-running interest in "how you can use existing causes of action effectively. How you can develop them, create them, move them along, and achieve social outcomes that are desirable?" Michael Rivette observed that there are some causes of action that "seem to have more flexibility built into them" and gave the example of breach of confidence: "in privacy litigation I will always look for a breach of confidence, because we've already developed that. It's a really powerful cause."

Indeed, litigators can actively work toward this kind of efficiency by strategically retrofitting doctrines over the course of several lawsuits for different clients. Rivette's approach to developing the law of breach of confidence over 14 years was, he said, conscious and targeted: "me and many others knitting the tapestry together". He spoke of a series of privacy and breach of confidence cases in which the remedies for privacy infringements expanded (*Giller v Procopets* 2008; *Jane Doe v Australian Broadcasting Corporation* 2007):

> it becomes public interest litigation to lower the threshold [for damages for breach of confidence] like we did in *Giller v Procopets*, to make that something that's more powerful in a setting where people's information and data is used to their detriment.

These cases changed the landscape by establishing that the equitable action for breach of confidence could apply to privacy harms, and that the threshold for damages for such a breach could be lowered to "mere distress", rather than more tangible harms.

## Bolt-ons and aligning incentives

A key dimension of retrofitting is aligning incentives. Public interest litigators are working in imperfect circumstances. Pragmatic litigators, we found, will cannily try to align private interests (litigants', funders and firms) with the public interest. To this end, litigation must do more than merely clarify the application of the law to new circumstances. The public interest dimension comes from seeking some outcome broader than the resolution of a particular dispute. We presumed there would be tensions between the public interest and the immediate interests of litigants in swift and cost-effective resolution. Our participants highlighted, however, that many litigants have the conviction to pursue more systematic change.

Participants were conscious not only of aligning claimants' incentives with the public interest, but also the incentives of other actors, including funders. Litigators pointed out that it was tactically sound to pursue, where possible, causes of action which provide monetary damages as a remedy for ADM harm. A public-interest-litigator-turned-legal-scholar, explained:

> These actions have to be profitable, so that litigation funders want to fund them, that's the bottom line, and yes, there's lots of arguments about what is the right percentage. And obviously you don't want it to be so much that people don't get proper damages for the harm, and the people that benefit are the lawyers and the litigation funders. But then again, you've got to incentivise law firms and litigation funders to bring those actions. Huge risk is involved, particularly for law firms.

In other words, it will make sense wherever possible to include in public interest claims causes of action for which *monetary remedies* are available.

Doctrines such as negligence, unjust enrichment, breach of confidence, and other torts and equitable actions, participants explained, are all appealing for that reason. The doctrine of negligence, for example, has three things going for it: the remedy is monetary; it has a history of incremental adaptation to new technology; and it is not necessary to advert to the mental state of the defendant (e.g. intent) in relation to the harm. On the other hand, as Carney pointed out, negligence may be difficult to establish in relation to government ADM, because courts are reluctant to recognise duties of care that interfere with existing statutory duties. Equitable doctrines may recognise and compensate for a wider range of injuries than tort, Rivette, explained. The recent reform to privacy law (allowing individual actions for serious invasion of privacy) may also generate incentives for litigators to focus on the use of data in ADM (Privacy and Other Legislation Amendment Act 2024).

Participants were generally not overly worried about involving commercial actors in public interest advocacy. They instead sought alignment between those actors' business interests and the public interest. For instance, Lizzie O'Shea pointed out that enterprising litigation funders may see the funding of a test case, even fairly creatively framed, as an investment in a future revenue stream: "I think funders would be willing to be a bit experimental with some of these kinds of things, because it's quite clearly a developing field". If an action is successful and generates a sufficiently important legal development, funders may then have the opportunity to fund (and profit from) future cases cut from the same cloth. The same logic applies to commercial firms, and even barristers. Actors with commercial interests, even interests in taking portions of damages awarded, were generally seen as enablers of effective action. That view is consistent with research which finds that US legal professionals' motivations for public interest work are varied, and mercenary motivations do not necessarily compromise public interest goals (Cummings and Sandefur 2013).

In formulating claims, remedies matter as much as *conceptual* and *doctrinal* potential. To address harms from ADM in government, Lizzie O'Shea pointed out that the "sweet spot" is the intersection between private and administrative law – for example in relation to government IT procurement. That view is consistent with work by McGurk and Tomlinson which emphasises the role of procurement law alongside administrative remedies in dealing with government ADM (McGurk and Tomlinson 2025). Incentives will be better aligned – especially for funders – in suits about government use of ADM when monetary remedies are available – rather than cases where only non-monetary administrative remedies such as declarations or injunctions are available. Robodebt litigation matched that pattern: a judicial review suit led government to stop using the Robodebt ADM system (*Amato* 2019), but it was the lawsuit in negligence and unjust enrichment, with its potential for large sums in damages, that made a class action viable, and (as Min Guo pointed out) motivated the government to compensate victims in the settlement (*Prygodicz v Commonwealth of Australia (No 2)* 2021).

Another key part of retrofitting includes what we call "bolt-ons" – adding broader public interest matters to otherwise private lawsuits. As Megan Richardson explained, an *amicus curiae* brief may permit a court to address public interest matters in the context of more narrowly framed lawsuits. She gave the example of the amicus brief by Michael Rivette for the well-known *Grubb* case (*Privacy Commissioner v Telstra* 2017). That case addressed the definition of personal information under the Privacy Act 1988. Richardson said:

> the Privacy Commissioner did have a public interest argument. It was just not fully developed as the argument that the *amicus* was making and keeps making. So I think there were just many different ways of talking about the public interest in that case.

In this instance, the amicus brief permitted the court to consider a wider range of public interest issues relating to the definition and regulation of personal information.

Michael Rivette described another tactic he used to expand the scope of lawsuits – which we also think of as a bolt-on tactic. He would include more creative and untested legal claims alongside more conventional, established claims. He explained:

> The first thing is how do I best protect the interests of my client? Number one; paramount. The second thing I think about is…what are the broader principles that can be developed in this case? That's the public interest thing. Now, provided the second doesn't work contrary to the first, then I will always develop that, always. And the reason I'll develop that is because if I'm putting the second to a court, and a court knocks me down and they've also knocked down the main case, I might get up on appeal on the second one. Because the court wants to hear it. And so that's often in the interest – if it's not against the interest of the client to develop the law as well.

This bolt-on tactic aligns the client's interest with the public interest. Adding a less established claim in effect diversifies the prospects of success. At the same time, it generates opportunities for systemic retrofitting that create a foothold for future suits. It also represents a practical pathway for private practitioners to realise a widely-held aspiration among lawyers: to use "ordinary" commercial practice to promote justice and positive systemic change (Sharpless 2012; Johanson 2020).

# Limits of public interest litigation

While the tactics described above are pragmatic, the fact that such pragmatism is needed reveals shortcomings of law and politics. Australian law is far better at dealing with concrete, individualised harms than what are often perceived as "intangible" harms (to mind, emotion and dignity), systemic harms (such as the unchecked expansion of government and corporate power and the normalisation of near-universal surveillance), and collective harms (such as algorithmically encoded discrimination). In exploring some of these disconnects below, interviews exposed deep structural problems in our legal system, its institutions and its legal cultures.

## Problematic laws, institutions and risk averse legal culture

Attempts to retrofit old laws for new circumstances frequently expose the problems with existing laws and institutions. Australian legal systems carry ongoing legacies of colonialism, racism, sexism, homophobia, and other injustices (Behrendt, Cunneen, and Libesman 2008). Even laws intended to redress some of these harms are limited. For instance, participants viewed anti-discrimination law, which might appear well adapted to address unfair ADM and algorithmic bias (Barocas and Selbst 2016), pessimistically.

Discrimination law is limited by a number of problems. Courts do not tend to award large sums in damages, so litigation funding lacks incentives. It lacks the flexibility of doctrines like negligence. And it is based on an outdated model that sees discrimination as an issue of individual bias or prejudice rather than a systemic structure. A litigator involved in movement work with Indigenous communities explained, memorably: "we find it easier to prove that a hospital's been medically negligent than they have been racist, even if the negligence was based on prejudice.” While negligence is unconcerned with the defendant’s state of mind, discrimination law has difficulties when discrimination arises from something other than bigoted intent.

Direct discrimination does not currently address algorithmic bias based on digital proxies – themselves derived from data reflecting historical injustice (Mann and Matzner 2019). In our interview, technology professor Lyria Bennet-Moses explained how existing law relies on outdated socio-technical assumptions:

> [I]n the case of the discrimination law, essentially, the assumption was bigoted humans. And there's bigoted humans who are explicit about being bigoted. Those cases are easy. There's bigoted humans where it's hard to tell and there are various ways to try and catch them out if it's very clear from the data, but not from explicit statements of bigotry. But this model is not a good fit for machines.

If intent is already difficult to prove, things become further complicated when ADM systems encode intersecting forms of discrimination (Crenshaw 2013). As tech law scholar, Jake Goldenfein, put it, “Your data is operationalised in a way that is not intelligible to humans”. There is considerable uncertainty about how the law applies to discrimination on the basis of an algorithmic profile, which is abstracted from protected characteristics such as sex and race (Wachter 2020).

The assumptions of a straightforward, intentional decision motivated by conscious animus no longer work with ADM. This problem is compounded by what Lauren Perry, a Policy and Projects Manager for Responsible Technology at the UTS Human Technology Institute, called “technology deference”:

> people go, oh, well, I guess the computer made that decision, and so that must be right. Computers and algorithms, they’re less biased than humans, and they’ve been coded to make the right decision, so we can’t really challenge that.

She recalled multiple instances where the person “on the other end of the phone was saying, well, you have to trust us, because… the system has told us x, y, z.” Automation bias (the tendency to assume that automation removes human “errors” such as discrimination) risks legitimising decisions that (literally) automatically reenact past injustice.

The doctrine of indirect discrimination seems a promising avenue for addressing automated inequality and injustice. But it too faces significant obstacles. In State and Federal law, generally, indirect discrimination occurs when a person imposes a requirement, condition or practice, that has, or is likely to have, the effect of disadvantaging persons with a protected attribute, and that is not reasonable (see e.g. Equal Opportunity Act 2010 (VIC), s 9(1)(a)). However, the standard of unreasonableness is uncertain with respect to automated decision-making. An academic and litigator specialising in discrimination reflected that judges have “really shied away from making judge-made law” in the area of indirect discrimination. So even where lawyers think the law could be applied, they may not take that avenue because they anticipate “there would be problems, in practice… because judges don’t want to assess if something’s reasonable or not, in this context.” This comment is consistent with scholarly analysis on the challenges of applying indirect discrimination to ADM decisions (Sheard 2022).

This points to a broader problem. Public interest litigation depends on legal culture – the attitudes of lawyers and judges to legal creativity and risk. Adapting the law to new technologies needs a legal profession prepared to apply the law flexibly.

Participants expressed mixed views about the risk appetite of Australian lawyers. Lizzie O'Shea mused about "whether you could convince a barrister to be experimental enough with their advice or be open-minded to thinking about" innovative retrofitting style-claims with respect to ADM. A public interest litigator, she decided, "I feel like there's enough resources at the bar. They're not all people who are averse to technology. There are people you can find who will work on these things with you, I think."

As for the bench, Jake Goldenfein worried that the Federal Court, with a "very black letter" orientation, might not be sufficiently open to new scholarly ideas regarding the development of privacy law:

> I wish there was somehow scope for ideas to proliferate their way through the arguments that get made in some of those cases. But it's hard because often these cases are about really narrow points [such as the definition of personal information]. It's very hard to bring a radical agenda about the state of data governance when it comes to something like that.

By contrast, Michael Rivette suggested that other doctrines, such as breach of confidence and negligence, seemed to be associated with a more flexible judicial culture; open to more scholarly or creative arguments, especially since the appointment of Justice Edelman – a former legal academic – to the High Court of Australia.[4] Taken together, participants' comments suggest that enough of the profession is curious and public-minded enough that they could be powerful advocates for clients harmed by ADM, but that community building strategies described above will play an important role in developing creative legal cultures.

## Heavy burdens of litigation limit access to justice

How the law develops in respect to ADM is further influenced by who can afford to bring legal actions – financially, practically and emotionally. Issues of concern to well-funded litigants will tend to dominate. As David Lindsay pointed out, privacy rights have expanded significantly through court action, especially in the UK, because of celebrity actions.

But people already experiencing marginalisation will face additional challenges in bringing public interest litigation, especially in the uncertain area of automated harms. As Samantha Floreani, then program lead at Digital Rights Watch (and now writing a PhD on "rent tech") put it,

> people who are getting harmed by [technology] are usually, generally, already vulnerable or marginalised or historically oppressed or continue to be oppressed. So, when that happens, they also get less attention…

The burdens of litigation may be especially heavy for people in this situation. Vicki Sentas summed up the problem in this way: "There's a lot of complexity to finding the right litigants and ensuring it's safe for them as well. And most of the time, it's not safe." Camilla Pandolfini expanded on the same sentiment:

> It feels very unfair that… it has to be individuals that bring these claims to try and bring about benefit for the community that they're part of… that they have to take on that individual risk themselves. Civil litigation is long and onerous and asks a lot of people, and one of those things is that they have to think about that adverse cost risk and the impacts on them.

[4] The High Court is Australia's highest court (as distinct from the High Court of England and Wales).

Litigation is extremely stressful, with huge time demands. As Lizzie O'Shea put it, "It takes over your life.". Participants further mentioned the risk to litigants of reputational harm including what Min Guo referred to as the airing of "dirty laundry": a particularly unappealing prospect for litigants who have already been impacted by the criminal justice system.

A partner at a major commercial law firm, with extensive experience in pro bono work felt that these burdens block access to justice for people already likely to be experiencing marginalisation. They were therefore very pessimistic about public interest litigation:

> If you had a dispute resolution mechanism that was accessible, cost-effective and genuinely resolved disputes instead of wearing people down as a process in itself, that would be great, but that's not what we've got.

Their statement evokes the adage "the process is the punishment", which describes how the experience of going through criminal proceedings in the US can be so taxing, hurried, and unfair as to be a punishment by itself (Feeley 1979). It also evokes feminist scholarship that draws attention to the ways sexual assault survivors are frequently re-traumatised by the criminal justice system (Graycar and Morgan 2002), and in particular how Indigenous women are disbelieved as witnesses (McQuire 2024) and punished for seeking assistance (Watego et al. 2021).

## Entrenching structural harms

Alongside rigid legal cultures and access barriers, there is a risk that public interest litigation cements broader structural harms, despite best intentions. Where litigation is embedded in a broader advocacy strategy around issues with political momentum, its prospects of success are generally better. This creates a subtle incentive to "fit within the dominant political paradigm" as David Lindsay put it. This may entail prioritising concepts like "safety" and promoting fear about new technologies, rather than taking the "purest view of law reform, which is to say what are the fundamental problems? How do we address those fundamental problems?"

Notwithstanding pragmatic efforts to align incentives, there will sometimes be irreconcilable misalignments between the immediate interests of a litigant and the public interest more generally. Bennet Moses put the problem in these terms:

> it's difficult to use litigation as a mechanism to fix systemic error, because you can always settle. And I think that it'll always be in any given litigant's interest, unless they're really committed to the public interest, which a lot of litigants in the end might not be when someone's saying, "I can solve your problem".

Participants were quick to point out that many public interest clients are deeply committed to broader social justice outcomes. Nonetheless, a client's interest in compensation and a speedy resolution of their complaint may be at odds with public interests in seeing a case through to judgment. In the worst case, Terry Carney pointed out, corporations and governments adopt a cynical strategy of separately settling individual cases – either in administrative review or in court – to obfuscate systemic problems and avoid judgments that produce change.

# Enablers for effective public interest litigation

What are the essential pre-conditions for public interest litigation about ADM to be effective? We identify three action points: transparency across accountability lifecycles, network-building, and funding.

## Transparency throughout the accountability lifecycle

The opacity of AI and ADM systems is well known. Scholars have pointed out different kinds of opacity – how algorithmic systems may be inscrutable, intentionally concealed and difficult for those without technical expertise to understand (Selbst and Barocas 2018; Burrell 2016; Fraser, Simcock, and Snoswell 2022). Our interviews added more colour to this picture. They revealed not only that there are different *kinds* of opacity that obstruct accountability with respect to ADM – but that there are different moments along a timeline of accountability where opacity (or transparency) matter in different ways. We gained this insight, which is not emphasised in existing literature, as a result of our participant selection. Stakeholders from civil society, academia and legal practice each contribute different stages of accountability-seeking with respect to ADM harms, and so could tell us about the role that transparency and opacity played in their own niche.

In the broadest sense, and throughout the whole lifecycle, opacity of and about ADM reinforces asymmetries of power. Subjects of unjust ADM may feel this viscerally. Lauren Perry, who had been undertaking qualitative sociological research into facial recognition technologies, and who was personally impacted by Robodebt, reflected that "for a lot of individuals, there's this real sense that new and emerging technologies are happening to them." In other words, opacity creates a sense of powerlessness.

Transparency about ADM is thus a pre-condition for accountability. Individuals, advocates, regulators and the general public all need various kinds of transparency at multiple points along the accountability timeline to perform their various accountability functions. Participants were generally in favour of horizontal AI regulation mandating, at a minimum, transparency requirements for ADM. Interviews took place before the minor tranche of reforms to the Privacy Act in late 2024, and also before the Department of Industry, Science and Resources proposed (and then abandoned) the introduction of "mandatory guardrails" for high-risk AI, which would have imposed basic documentation, transparency, testing and risk management requirements (*Safe and Responsible AI in Australia* 2024). As Kimberlee Weatherall has put it: this kind of regulation would create a "paper trail" that could then be used to enforce other laws (Weatherall 2025).

Aside from direct explanations about ADM to users, participants reminded us that various kinds and degrees of transparency could provide a foundation for public interest legal action. Kate Bower, working at the time of our interview at CHOICE, a consumer advocacy organisation, explained the value of transparency in facilitating accountability through "peer review":

> I think there's a level of transparency that's needed for consumers. But I actually think there's a big role in terms of accountability, those people I was talking before, researchers, regulators… journalists all have a role to play.

Without this kind of transparency about ADM deployment and function, it may be difficult or impossible even to detect ADM harm or understand its extent. Sam Floreani took the view that transparency measures such as documentation of ADM design, and inspectability of algorithms would "enable [civil society] organisations like Digital Rights Watch to have an increased ability to see what's going on and then respond or help people raise awareness." As a discrimination scholar pointed out, this kind of transparency – mediated by journalists, academics or public interest organisations – has the capacity to "raise community awareness about where these algorithms might be used, so they know where they might be impacted." In short, transparency can operate on many levels. It can be directed at multiple audiences who might all play different roles in surfacing problems, and progressing them toward appropriate accountability mechanisms, including litigation.

Each layer of transparency then facilitates further transparency and accountability mechanisms. Once there is sufficient transparency to identify a pattern of wrongdoing, to bring it to a lawyer, evaluate prospects of success and bring a claim (all formidable challenges), litigation itself can shed further light on ADM harms. This was Vicki Sentas' view: "I think its primary modest benefit is the exposure of data, which illustrates the harm." On this view, litigation is itself a mechanism for transparency through its process of discovery and evidence gathering – and it is this role in bringing to light injustice which permits it to work effectively as part of an overall ecosystem of accountability and advocacy.

For litigation to work, however, lawyers need to understand wrongs and explain them satisfactorily to courts. Technical capability in the profession is therefore a condition for effective litigation. As Ed Santow put it, "there's really important upskilling that needs to happen throughout the legal profession". But Lizzie O'Shea reminded us that judges, despite generally coming from an older generation that is perceived to be less tech savvy, are already heavily engaged with technology:

> Judges routinely now have to approve plans for production of documents. That involves teaching a machine to identify relevant material, putting in place processes to correct errors, and then using that as a way to streamline discovery so it doesn't totally bog down a case… there are aspects of practice already that judges have to come to terms with sophisticated technology.

Min Guo suggested that enhanced technical capability in the profession might be a question of time, exposure and familiarity. Since our interviews, the use of AI-based technologies has certainly become increasingly prevalent throughout the profession (Bell 2025).

Technical capabilities around ADM could further be strengthened by involving expert witnesses. One human rights lawyer reflected, "there is good technical expertise out there… I'm sure you'd get expert evidence". Lizzie O'Shea further explained,

> None of this is particularly difficult, but you're probably going to have to analyse code. You're going to have to have an expert who looks into the machine learning aspect and tells you how the discrimination came to pass. So, it's not like your standard litigation.

To facilitate such upskilling, and to support a robust transparency-accountability lifecycle, the aggregation and sharing of information among stakeholders is essential.

## Networks for information sharing and community mobilisation

Transparency at various points in the accountability lifecycle can only lead to meaningful change if stakeholders are able to join the dots. Currently, the ecosystem for recognising harm and progressing complaints is what one human rights lawyer called "fragmented", with civil society organisations and regulators working in what consumer advocate Chandni Gupta called "silos". Terry Carney talked about how systemic issues may slip through the gaps:

> [a] myriad of small issues that an administration can get away with messing up or acting illegally because nobody has, it's often not finance, but just the mental stamina to think that it's worth their while. It appears too trivial.

Participants had a range of suggestions about how to build what we came to think of as *information aggregating mechanisms*. One common thread was the urgent need for effective mechanisms for receiving, aggregating and analysing complaints to detect patterns of systemic wrongdoing. David Lindsay put it this way: "Litigation is the tip of the iceberg. It's all of the complaints that people have day after day of navigating systems in which they feel disempowered." A social justice litigator elaborated on how accessible complaint-making is a key condition for ADM accountability:

> we have a theory of change that online complaint-making (particularly about discrimination) can make a difference. And the data that's collected from complaint-making can support evidence-based advocacy to drive social change. So we're collecting information on discrimination and incidents of racial hatred. In time we will build more complaint pathways. And we're also designing an effective technological pathway to making complaints easy.

Lily Ryan, an information security specialist at Thoughtworks, noted that this could prompt further investigations through mechanisms such as freedom of information requests or inspections of code and algorithms.

Another mechanism for aggregating and distributing information is through appropriate allocation of regulatory responsibility. Terry Carney spoke of the need for institutions capable of systematically reviewing administrative complaints and processes, and argued for the reinstatement of Australia's Administrative Review Council, which was composed of experienced decision-makers "much more attuned" to identifying administrative issues at a systemic level. Since the time of interviews, that body has in fact been reconstituted (Administrative Review Tribunal Act 2024).

Formal institutions with responsibility for systematic monitoring and review could therefore complement public interest litigation. Such institutions could modulate the problem that individual proceedings (whether for administrative review or judicial review) may be piecemeal and, on their own, inadequate to meet systemic harm. Regulators too have a role, as Chandni Gupta, Digital Policy Director at the Consumer Policy Research Centre, said:

> what we really need to see regulators especially move towards is a proactive surveillance approach. Them building their capacity and capability within their workforce, of a very diverse workforce that's got the technical knowhow, and not just

> the legal knowhow, to really do the deep dives into how are these systems working. And be able to identify harm before mass harm occurs.

Australia's new National AI Plan established an AI Safety Institute to guide the development of safety practices and guardrails (*National AI Plan* 2025). While its remit seems at present to be highly technical – developing technical practices and guardrails to promote "safety" – there is an opportunity for the Institute to expand its impact by helping regulators in multiple domains develop better expertise and practices for identifying and responding to ADM harm at scale.

It is not enough, though, just to *know* about problems at a systemic level, or to rely on a review body or regulator to correct them. Part of the function of public interest litigation is as a "private regulatory tool" (O'Shea). No one knows the impact of digital harms better than those who suffer them. Community organisations have enormous expertise. Those who are affected by ADM also need to know how they can take action, and who will assist them. Part of this includes promoting legal literacy within communities. As Lauren Perry pointed out "ways to increase community awareness around access to justice" is essential. This kind of network building would also be an important path to address power dynamics whereby lawyers and those with specialist technical knowledge, rather than those more directly affected by injustice, come to dominate social movements (Handler, Hollingsworth, and Erlanger 1978).

There is a foundational need for better solidarity networks and long-term relationship building, connecting would-be litigants, community organisations and consumer bodies with the relevant resources and people. One human rights lawyer spoke about

> a lack of an interconnectedness within the legal system that connects people who are suffering with people who have the ability to bring cases to address that suffering. Unfortunately, again there's resourcing constraints at every level and so that sort of joined-up ecosystem, one reason it's not fully integrated is a lack of capacity to play. Everyone's so busy trying to do their job that that higher level dimension goes missing.

Participants proposed various options to address this problem, including a specialist digital legal centre, and the development of effective referral networks. Ed Santow pointed out that the kinds of "hackathons" described above were an effective, informal way of developing these kinds of networks, from which collaboration, information sharing and referrals could develop organically: "it doesn't need to have a lot of structure".

## Funding support for litigation and litigants

For public interest litigation to elicit transparency and accountability, it requires funding and support – as well as the removal of unnecessary cost-barriers. Samantha Floreani and Jake Goldenfein both pointed out how relatively "immature" Australia's civil society landscape was, compared to the United States and the UK. Organisations that might aggregate information and coordinate action lack funding – which may be why our participants tended to see lawyers as key drivers of litigation.

At later stages in the accountability lifecycle, litigation cost support is also crucial. Community legal centres and Legal Aid need funding to smooth the transition from grievance to legal action. Min Guo pointed out that Legal Aid would require "endless resources", and made the more radical proposal of developing a universal public legal services system, like Medicare. Other participants emphasised the need to ensure that private litigation funding remains viable.

But support must take forms other than direct funding. Many participants pointed out how obstructive the risk of adverse cost orders is to a would-be litigant. One serious problem is the fact that lead plaintiffs in class actions are solely liable for adverse cost orders (whereas any proceeds of the case are shared with the whole class). In one human rights lawyer's words, "adverse costs orders against your lead plaintiff can stymy litigation or deter people from litigating". A more generous arrangement regarding cost orders would significantly change the landscape – a view held by many of the litigators in our study. In the Robodebt case, it was the plaintiff law firm, Gordon Legal, that indemnified applicants against adverse costs orders, as well as acting on a no-win-no-fee basis. But a more systematic solution is warranted. While indemnities may help, getting them may be time consuming and difficult.

Public interest exceptions to such orders would be one such option. Another option would be to provide that in public interest litigation against government, only the government should have to pay adverse costs. Ed Santow pointed out that other jurisdictions have mechanisms of this kind, such as the protective costs orders available in public interest litigation in the UK. These arrangements have not, despite their critics' concerns, opened the floodgates (Lock QC and Mills 2016). All of the other burdens of litigation discussed here provide a strong disincentive against frivolous litigation. The UK might therefore provide a template for reform to Australia's adverse cost regime.

On that note, for public interest litigation to be truly effective, litigants would also need support in meeting non-financial costs. Lizzie O'Shea explained that to properly support litigants under the extreme strain of litigation would require "a whole different social infrastructure and organisations to assist and work with them." It makes sense to think of funding and support of accountability mechanisms in terms of ecosystems and lifecycles: not simply funding one action, but sustaining a whole infrastructure. A better-funded civil society, could develop the structures, expertise and systems to properly support litigants across the accountability lifecycle.

# Conclusion

Increasingly, automated decision-making is being used in an enormous variety of areas: in the provision of social services, predictive policing, rental applications, employment applications, financial services, and in health and education systems. Digital proxies (which we may never know about) are being used to determine our access to basic rights and services. These decisions are not "intelligent" or even conscious assessments; they are guesstimates and approximations, built on flawed datasets and problematic assumptions. Many systems are opaque, unaccountable, and operate with near impunity. Emboldened by geo-political pressure to reduce regulation of big tech, and fuelled by corporate greed or government fetishisation of "efficiency", they continue to avoid accountability, despite whistle-blowers, leaked documents and investigative journalism exposing their harmful effects.

In a milieu where the prospects of effective ADM regulation are uncertain, it is essential to cultivate the conditions necessary for bold and ambitious public interest litigation. By this we mean litigation that not only asserts or develops the law, but also supports broader social movements for justice. According to our participants, the strategies of public interest litigation for automated harms are multiple and overlapping. They include reminding ADM developers and deployers that they are not above the law; and exposing gaps in the law to precipitate law reform. At times, this involves losing to win. A central strategy of public interest litigation is to be in service of a larger public interest agenda.

Public interest litigators work in an imperfect environment that demands ingenuity. They described various tactics: retrofitting old laws to new harms, bolting-on public interest arguments to private suits, aligning private and public interests where possible. Absent a Federal bill of rights, it is prudent for activist communities to "hack" together to build "theories of the case" so that they are ready to take action when the right litigant appears. Our pragmatic participants made it clear that public interest litigation is most likely to succeed when it addresses harms to which our law and legal institutions are best fitted: individualised harms with damages as a remedy.

Accountability has a lifecycle. Different kinds of transparency matter to different stakeholders at different stages of this lifecycle – from detecting harm in the first place, to obtaining a judgment in court. While public interest litigation is often seen as the last stop in the lifecycle, it is also generative. It exposes mechanisms, unearths data, produces media, builds momentum, and plants seeds for further change.

But even with better transparency, effective networks will be essential to progress the detection of ADM wrongs through the accountability lifecycle. Legal retrofitting only works if the profession, bar and bench develop and maintain a *culture of progress:* cultivating the technical expertise and the courage to develop the law creatively for new circumstances and towards public interest goals. We therefore need to build relationships and cultivate networks for referrals, information sharing, and theory development.

All of this requires energy, funding and support. Litigants need financial support, but also support in dealing with the personal strain of litigation. Establishing a system in which individuals or classes can bring public interest litigation against corporations and governments without adverse costs could make inroads on the massive power imbalance between these parties. All of these enablers are mutually reinforcing – a more conducive funding landscape, more accessible information, and more robust networks will support a stronger culture of activism and advocacy for justice in the age of digital automation.

These findings also reveal the need for broader law reform to *prevent* harm from automated decision-making. Public interest litigation is a last resort to address harm that has already occurred. Uses of ADM that will always be unjust (such as social scoring or negative targeting based on disability) should be prohibited, and regulation should require that the design of automated systems meets human rights standards and works to redress inequities rather than reproduce them. This is something more than simply requiring systems to be "unbiased." Automated decision-making should be used in ways that make access to service and opportunity for marginalised communities easier, rather than harder (Pasquale 2022).

At a minimum, ADM providers and deployers should be obliged to meet basic standards of transparency. Transparency requirements should be savvy to the systems of opacity and obfuscation that obstruct accountability– from complaints to litigation – and should build systems for contestability, including mechanisms for reparations where automated systems cause harm or breach human rights. Australia had a sound proposal, backed by a rigorous consultation process, to introduce "mandatory guardrails" for high-impact AI and ADM, including significant transparency mechanisms (*Safe and Responsible AI in Australia* 2024).

Perhaps it is naive to recommend the government revive that proposal, but abandoning it – with little explanation – was a mistake. Broader regulatory changes, including formal recognition of collective, structural and indirect harms are also needed, starting with reform to make the law of indirect discrimination more workable. Without these minimum protections, the risk is that, even after public interest litigation, harmful automated systems are just adjusted and continue in new forms.

There are some promising developments, however. The Australian government is considering a "digital duty of care" that could apply to AI and ADM providers (Department of Infrastructure, Transport, Regional Development, Communications, Sport and the Arts 2025), alongside regulation that would apply to government use of ADM similar to the approach taken in Canada (Attorney-General's Department 2025). We welcome both. Regulation of government use of ADM will increase transparency, while a digital duty of care could create opportunities for public interest litigation by removing the hurdle of establishing a duty under negligence or other doctrines.

In the meanwhile, and even after regulation arrives, we must make the best possible use of all the tools we have. We hope to have done justice to our participants' vision of an accessible, inclusive ecosystem for public interest litigation, nested in strategies for advocacy, driven by affected communities who are experts in their own experience, and supported by a canny, innovative and justice-oriented legal profession.

# Cases cited